\documentclass[12pt]{iopart}
\usepackage{graphicx} 
\usepackage{amsfonts}
\begin{document}

\title[Light trapping within diffraction gratings]{Light trapping within the grooves of 1D diffraction gratings under monochromatic and sunlight illumination.}

\author{Mario~M.~Jakas$^\dag$ and Francisco~Llopis$^\ddag$}

\address{Departamento de F\'{\i}sica Fundamental y Experimental, Electr\'{o}nica y Sistemas. Universidad de La Laguna, 38205 La Laguna, Tenerife, Spain} \ead{$^\dag$mmateo@ull.es  $^\ddag$fllopis@ull.es} 

\begin{abstract}
The Rayleigh-Modal method is used to calculate the electromagnetic field within the grooves of a perfectly conducting, rectangular-shaped 1D diffraction grating. An \emph{enhancement coefficient} ($\eta$) is introduced in order to quantify such an energy concentration. Accordingly, $\eta >$1 means that the amount of electromagnetic energy present within the grooves is larger than that one will have, over the same volume, if the diffraction grating is replaced by a perfectly reflecting  mirror. The results in this paper show that $\eta$ can be as large as several decades at certain, often narrow, ranges of wavelengths. However, it reduces to approximately 20\% under sunlight illumination. In this latter case, such values are achieved when the \textit{optical spacing} between the grooves $dn$ is greater than 500 nm, where $d$ is the groove spacing and $n$ is the refractive index of the substance within the grooves. For $dn$ smaller than 500 nm the enhancement coefficient turns negligibly small. 
\end{abstract}



\maketitle

\section{Introduction}
\label{Introduction}

It is well documented that electromagnetic fields within the grooves of diffraction grating can be largely increased when illuminated by light \cite{Glass83, Wirgin85, Lochbihler93, Depine94}. This phenomenon was recently proposed as way to enhance the absorption of light in PV cells and optoelectronic devices \cite{Crouse05,Lee08,Llopis10, Wang10}. In fact, for the purpose of increasing the efficiency of photo-cells where sunlight may not be absorbed so easily, the \emph{field enhancement} appears to be a useful approach. In this realm, however, rather than differential one needs integrated figures. This is so because large field enhancements, as those previously reported, are not enough if not accompanied by a net increase of the light energy integrated over a representative volume of the cell, along the pertinent range of wavelengths, incidence angles and states of polarization. 

In a previous paper of the Authors \cite{PIERS2011}, the field enhancement was analysed for the more general case of beams of monochromatic light arriving to the grating along several directions and different states of polarization. There, an \emph{enhancement coefficient}, namely $\eta$, is introduced, so that $\eta >$1 means that the amount of electromagnetic energy present within the grooves is larger than that one will have, over the same volume, if the diffraction grating is replaced by a perfectly reflecting  mirror. In this paper, however, the previous study is extended to the case of illuminating the grating with solar light.  The results of these calculations show that, when using monochromatic, polarized and well-collimated beams of light, $\eta$ exhibits a series of peaks at certain wavelengths, where the enhancement coefficient can be as large as several decades, or even grater. However, after taking an average over incidence angles and polarization states, $\eta$ is significantly reduced as it may reach then values which can hardly be larger than approximately 3. If the previous results are also averaged over the solar spectrum, the enhancement coefficient is further reduced to approximately 1.20. 

Although these results are to some extent discouraging, it does not mean that the diffraction grating structure cannot be advantageously used in designing photo-electronic devices. This is so because, depending on the case, the aforementioned 20\% gain may suffice. In this regard, the present study could be interesting to those who may need increase the absorption of light within a solar cell and might possibly be planning to incorporate a diffraction grating for such a purpose. This paper is organized as follows: the basics equations necessary to obtain the field within the grooves of a diffraction grating are derived in Section \ref{Basic equations}. The results of numerically calculating the enhancement coefficient as a function of the various parameters in the model are shown and discussed in Section \ref{Results} and, finally, Section \ref{Summary} contains a summary and concluding remarks.

\section{Basic equations}
\label{Basic equations}

\begin{figure}
\centering
\includegraphics[width=8.3cm]{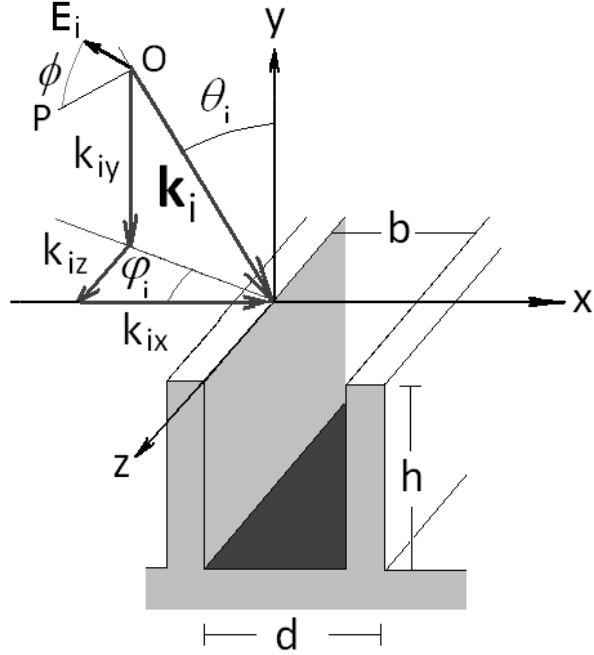}
\caption{A linearly polarized electromagnetic wave, with wave vector $\mathbf{k}_i$, electric field $\mathbf{E}_i$, incidence angle $\theta_i$ and azimuthal angle $\varphi_i$, arrives from the upper half-space upon a rectangular-shaped diffraction grating. Grooves have a spatial period $d$, width $b$ and depth $h$, and $\phi$ denotes the polarization angle with respect to the (x-z)-plane.} 
\label{Fig_1}
\end{figure}

The calculation model is sketched in figure \ref{Fig_1}. There, one can observe the incident light, a linearly polarized plane-wave, arriving to the grating from above at an angle $\theta_i$ with respect to the $y$-axis and azimuthal angle $\varphi_i$ with respect to the x-axis. The light has a wavevector $\mathbf{k}_i = (k_{ix},k_{iy},k_{iz})$, which is related to the wavelength in the upper medium ($\lambda$) as $\left| \mathbf{k}_i \right| = k_i = 2 \pi/ \lambda$. The electric field amplitude $\mathbf{E}_i$ is assumed to be perpendicular to $\mathbf{k}_i$ and $\phi$ is the angle formed by $\mathbf{E}_i$ and the ($y=0$)-plane. 

The diffraction grating is assumed to be an infinite set of equally spaced, rectangular-shaped grooves made on the surface of a perfect conductor. The spatial period of the grooves is $d$, whereas the width and depth are $b$ and $h$, respectively. Finally, the upper medium is assumed to be a non-lossy dielectric material that fills the upper part of the grating, up to a front surface which will not be taken into account in this calculations.

The way the electromagnetic fields within the grooves are calculated was already shown in Ref.\cite{Llopis10,PIERS2011}, however, for the purpose of the present analysis, it is important to re-do it once more. 

In the first place, the \emph{Rayleigh expansion} is assumed for upper space, 
 
\begin{equation}
\label{Eq1}
\mathbf{E}^{(y>0)}(\mathbf{r}) = \mathbf{E}_i e^{i\mathbf{k}_i\cdot\mathbf{r}} + \sum_{n} \mathbf{R}_{n}\, e^{i\mathbf{q}_n\cdot\mathbf{r}} \hspace{0.1\columnwidth} \mbox{for $y>0$} \, , 
\end{equation}

\noindent where $\mathbf{q}_n$ is the wave vector of the $n$-th order reflected beam and $\mathbf{R}_n=(R_{nx},R_{ny},R_{nz})$ is the corresponding electric field vector. Similarly, $\mathbf{q}_n = (q_{nx},q_{ny},k_{iz})$, where $q_{xn} = k_{ix}+2\pi n/d$, $n=0,\pm 1,\pm 2, \ldots$, and $q_{ny} = +\sqrt{k_i^2-q_{nx}^2-k_{iz}^2}$. Notice that the plus sign in front of the square root means that the positive, either real or imaginary solution to the square root has to be used. 

Within the grooves one has the so-called \emph{modal waves}, namely,

\begin{eqnarray}
E_x^{(y<0)}(\mathbf{r}) = & \displaystyle\sum_{m,N}{ {\cal E}_{x,m}\,  e^{i\left(k_{iz}z+Nk_{ix}x\right)}\,X_{m,N}^{'}(x)\,Y_{m}(y)} \nonumber \\
E_y^{(y<0)}(\mathbf{r}) = & \displaystyle\sum_{m,N}{ {\cal E}_{y,m}\, e^{i\left(k_{iz}z+Nk_{ix}x\right)}\,X_{m,N}(x)\,Y_{m}^{'}(y)} \\
E_z^{(y<0)}(\mathbf{r}) = & i\,\displaystyle\sum_{m,N}{ {\cal E}_{z,m}\, e^{i\left(k_{iz}z+Nk_{ix}x\right)}\,X_{m,N}(x)\,Y_{m}(y)} \nonumber 
\label{Eq2}
\end{eqnarray}

\noindent where 

\begin{eqnarray}
X_{m,N}(x) = \left\{
   \begin{array}{cl} \sin{\left[K_{mx}(x-x_N)\right]} & \mbox{for $x_N < x < x_n+b$} \\
   0 & \mbox{otherwise} 
   \end{array} \right. &
   \nonumber \\
X_{m,N}^{'}(x) = \left\{
   \begin{array}{cl} \cos{\left[K_{mx}(x-x_N)\right]} & \mbox{for $x_N < x < x_N+b$} \\
   0 & \mbox{otherwise} 
   \end{array} \right. & \\
Y_{m}(y) = \sin{\left[K_{my}(y + h)\right]}  \mbox{~~~~and~~~~} Y_{m}^{'}(y) = \cos{\left[K_{my}(y + h)\right]}\, , &  \nonumber 
\label{Eq3}
\end{eqnarray}

\noindent  where $K_{mx} = m\pi/b$ for $m=0,1,2,\ldots$ and $K_{my} = + \sqrt{ k_i^2 - k_{iz}^2 - K_{mx}^2}$, where, again, the positive solution to the square root must be taken. Similarly, $x_N$ denotes the $x$-coordinate of the right wall in the $N$-th groove, i.e. $x_N=Nd$ for $N = 0,\pm1,\pm2,\ldots$.  

Both, the continuity of the $x$ and $z$ component of the electric field along the groove aperture, and the boundary conditions of electromagnetic fields on the surface of a perfect conductor (see Refs.\cite{Kong86, Jackson}), allow one to write

\begin{eqnarray}
\label{Eq5a}
E_{ix}\,e^{ik_{ix}x} + \displaystyle\sum_{n=-\infty}^{+\infty}{R_{nx}\,e^{iq_{nx}x}} =  
\left\{
\begin{array}{ll} 
\displaystyle\sum_{m=0}^{+\infty}{{\cal E}_{x,m}\,\,\cos{\left(K_{mx}x\right)} \,  \sin{\left(K_{my}h\right)}} & \mbox{$0 < x < b$} \nonumber \\
0 & \mbox{$b \leq x < d$} 
\end{array} \right. \\
\end{eqnarray}
%
\noindent and, 
\begin{eqnarray}
\label{Eq5b}
E_{iz}\,e^{ik_{ix}x} + \displaystyle\sum_{n=-\infty}^{+\infty}{R_{nz}\,e^{iq_{nx}x}} =  
\left\{
\begin{array}{ll} 
i \displaystyle\sum_{m=0}^{+\infty}{{\cal E}_{z,m}\,\,\sin{\left(K_{mx}x\right)} \,  \sin{\left(K_{my}h\right)}} & \mbox{$0 < x < b$} \nonumber \\
0 & \mbox{$b \leq x < d$} 
\end{array} \right. \\
\end{eqnarray}

\noindent By taking the Fourier transform of these equations in the $x$ variable and after rearranging terms, one obtains

\begin{eqnarray}
\label{Eq6a}
R_{nx} = - E_{ix}\,\delta_{n,0} + \frac{1}{d} \displaystyle\sum_{m=0}^{+\infty}{{\cal E}_{x,m}\,\,\widetilde{C}_{m,n}  \,  \sin{\left(K_{my}h\right)}}
\\
\label{Eq6b}
R_{nz} = - E_{iz}\,\delta_{n,0} +  \frac{i}{d} \displaystyle\sum_{m=0}^{+\infty}{{\cal E}_{z,m}\,\,\widetilde{S}_{m,n}  \,  \sin{\left(K_{my}h\right)}} 
\end{eqnarray}

\noindent where,

\begin{equation}
\label{Eq7}	
\widetilde{S}_{m,n} = \int_{0}^{b} {dx\,e^{-iq_{nx}x}\,\sin{\left(K_{mx}x\right)}} \mbox{~~~and~~~} \widetilde{C}_{m,n} = \int_{0}^{b} {dx\,e^{-iq_{nx}x}\,\cos{\left(K_{mx}x\right)}}\, . \,   
\end{equation}

Similarly, the continuity of the $x$ and $y$ components of the magnetic field along the grooves aperture yields

\begin{eqnarray}
\label{Eq8a}
\lefteqn {2\,\frac{\left(k_{iy}^2+k_{iz}^2\right)\,E_{iz} + k_{ix}k_{iz}E_{ix}}{k_{iy}}\,\exp{\left(ik_{ix}x\right)} = } ~~~~~~~~~~~~~~~~~~~
\nonumber \\
& & \displaystyle\sum_{m=0}^{+\infty} \left\{ \left[ \frac{K_{mx}k_{iz}}{K_{my}} \sin{\left(K_{mx}x\right)} \cos{\left(K_{my}h\right)} -  \sin{\left(K_{my}h\right)}\, U^{(xx)}_m(x)\right]{\cal E}_{x,m} \right. \nonumber \\
& & + \left. \left[ \frac{K_{my}^2+k_{iz}^2}{K_{my}} \sin{\left(K_{mx}x\right)} \cos{\left(K_{my}h\right)} -  \sin{\left(K_{my}h\right)}\, U^{(xz)}_m(x)\right]{\cal E}_{z,m} \right\}\, , \nonumber \\
\end{eqnarray}

\noindent and,

\begin{eqnarray}
\label{Eq8b}
\lefteqn {2\,\frac{\left(k_{iy}^2+k_{ix}^2\right)\,E_{ix} + k_{ix}k_{iz}E_{iz}}{k_{iy}}\,\exp{\left(ik_{ix}x\right)} = } ~~~~~~~~~~~~~~~~~~~ \nonumber \\
& & \displaystyle\sum_{m=0}^{+\infty} \left\{ \left[ \frac{K_{mx}^2+K_{my}^2}{K_{my}} \cos{\left(K_{mx}x\right)} \cos{\left(K_{my}h\right)} -  \sin{\left(K_{my}h\right)}\, U^{(zx)}_m(x)\right]{\cal E}_{x,m} \right.  \nonumber \\
& & + \left. \left[ \frac{K_{mx}k_{iz}}{K_{my}} \cos{\left(K_{mx}x\right)} \cos{\left(K_{my}h\right)} +  \sin{\left(K_{my}h\right)}\, U^{(zz)}_m(x)\right]{\cal E}_{z,m} \right\}\, , \nonumber  \\
\end{eqnarray}

\noindent where 

\begin{eqnarray}
U_m^{(xx)}(x) =   \frac{1}{d}\displaystyle\sum_{n=-\infty}^{+\infty}{\frac{q_{nx}k_{iz}}{q_{ny}}\,\widetilde{C}_{m,n}\,e^{iq_{nx}x}}\, , & & U_m^{(xz)}(x) =  \frac{i}{d}\displaystyle\sum_{n=-\infty}^{+\infty}{\frac{q_{ny}^2+k_{iz}^2}{q_{ny}}\,\widetilde{S}_{m,n}\,e^{iq_{nx}x}}\nonumber \\ 
& & \\
U_m^{(zx)}(x) =  \frac{i}{d}\displaystyle\sum_{n=-\infty}^{+\infty}{\frac{q_{nx}^2+q_{ny}^2}{q_{ny}}\,\widetilde{C}_{m,n}\,e^{iq_{nx}x}}\, , & & ~~ U_m^{(zz)}(x) =  \frac{1}{d}\displaystyle\sum_{n=-\infty}^{+\infty}{\frac{q_{nx}k_{iz}}{q_{ny}}\,\widetilde{S}_{m,n}\,e^{iq_{nx}x}}  \nonumber
\label{Eq9}	
\end{eqnarray}

It must be noticed that no equations for the $y$-components of the field are necessary since the null divergence condition, i.e. $\nabla\cdot \mathbf{E}$ = 0, implies that only two, out of the three components of both $\mathbf{E}^{(y>0)}$ and $\mathbf{E}^{(y<0)}$ are independent. Finally, the values of ${\cal E}_{x,m}$ and ${\cal E}_{z,m}$ can be found by evaluating Eqs. (\ref{Eq8a}-\ref{Eq8b}) over a finite set of equally spaced $x$'s along the groove aperture, and the resulting system of linear algebraic equations is then solved by resorting to the Gauss elimination algorithm in Ref.\cite{NUMREC}. Once ${\cal E}_{x,m}$ and ${\cal E}_{z,m}$ are obtained, the reflection amplitudes, i.e. $R_{nx}$ and $R_{nz}$, can readily calculated from Eqs.(\ref{Eq6a}-\ref{Eq6b}).

Having arrived at this point, the \textit{enhancement  coefficient} ($\eta$) is introduced, namely,

\begin{equation}
\eta = \frac{1}{2\,A\,|\mathbf{E}_{i}|^2 } \int_0^b dx \int_{-h}^0 dy\,  |\mathbf{E}(x,y)|^2\, ,
\label{Eq10}	
\end{equation}

\noindent where $|\mathbf{E}_i|$ is the electric field amplitude of the incoming light, $\mathbf{E}(x,y)$ is the electric field within the groove and $A$ is the cross sectional area of the groove, i.e. $A=hb$. Notice that no integration over the $z$-coordinate is necessary since, for 1D gratings $|\mathbf{E}(x,y)|^2$ does not depend on $z$. 

As was previously mentioned \cite{PIERS2011}, such a coefficient allows one not only to quantify the field enhancement but also, and most important, it tells when there is a net gain of electromagnetic energy within the grooves. Actually, one can readily see that $\eta$ = 1 denotes the case for which, on an average, the electrical energy density within the grooves is the same as that present above, far away from the diffraction grating. 

\begin{figure}[ht]
\includegraphics{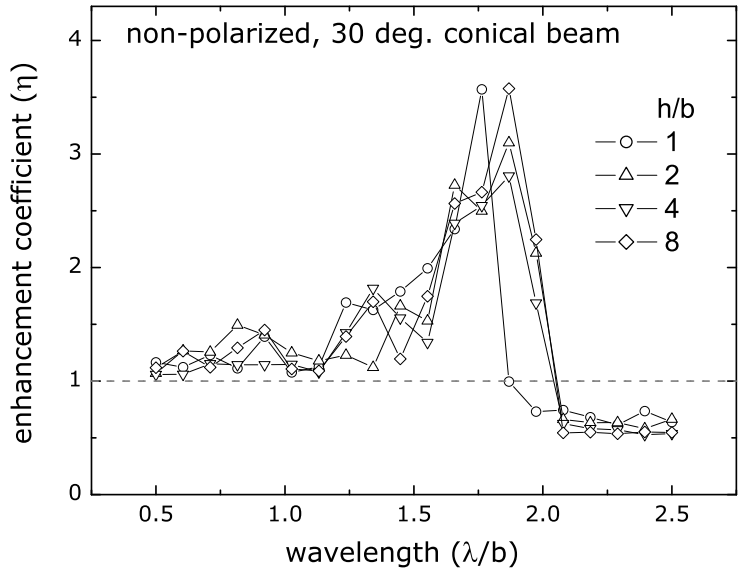}
\centering
\caption{Enhancement coefficient for a $h/b$=1 ($\bigcirc$), 2 ($\bigtriangleup$), 4 ($\bigtriangledown$) and 8 ($\Diamond$) diffraction grating, illuminated by non-polarized light, uniformly distributed over a 30-degree-wide, vertical cone.}
\label{Fig_2}
\end{figure}

It must stressed that the main goal in this paper is the calculation of $\eta$  as well as analysing its sensitivity upon the various parameters of the model. Before going into the results, however, it must be noticed that for the purpose of exploiting the trapping of light within the grooves, one can reasonably assume that $d=b$. By doing so, the number of variables is reduced by one and $b$ can be used as the unit of length. Accordingly, apart from $\theta_i$, $\varphi_i$, $\left|\mathbf{E}_i\right|$ and  $\phi$, the solutions to Eqs.(\ref{Eq8a}-\ref{Eq8b}) must be functions of $bk_i$ (or $\lambda/b$), and $h/b$. The results of numerically calculating these equations under the aforementioned assumptions are produced in the following section. It must be mentioned however, that all along the following sections, when referring to wavelength it is assumed to be that in vacuum. Only within the numerical code,  the wavelength is translated into that of Silicon using the refractive index from Ref. \cite{Phillip60}.

\section{Results and discussions}
\label{Results}

The enhancement coefficient is calculated for a 30-degree conical beam of non-polarized light. The results are plotted in figure \ref{Fig_2}, where $\eta$ appears as a function of the wavelength and for grooves of depths $h/b$=1, 2, 4 and 8. It must be mentioned that these results are obtained by calculating the mean-value of $\eta$ in Eq.(\ref{Eq10}) over the pertinent range of polarization states, incidence angles and wavelengths. This is performed by means of a Monte-Carlo integration scheme, where relative uncertainties of the order of ten percent, or less, is normally achieved. 

As one can readily see in figure \ref{Fig_2}, the enhancement coefficient exhibits a single large peak around $\lambda/b \approx 1.8$, becoming as large as approximately 3 for the four groove depths analysed in this paper. For $\lambda/b\leq$1.8, $\eta$ decreases and becomes slightly greater than unity for $\lambda/b \leq$ 1.2. However, one can see that $\eta$ falls below unity as soon as $\lambda/b>$2. Curiously enough, the enhancement coefficient does not exhibit a strong dependence with the groove depth $h$ and, as a matter of fact, differences between the results corresponding to the different groove depths are masked by the noisy aspect of the curves. This noise stems in part from the statistical fluctuations in the Monte-Carlo calculations, and also from the narrow resonances occurring within the groove which, as will be seen below, have been considerably reduced as a result of taking averaged values.

\begin{figure}[ht]
\includegraphics{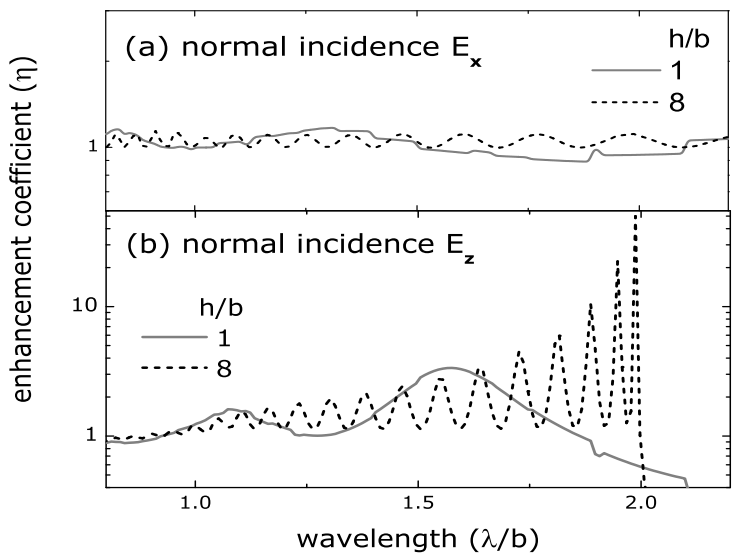}
\centering
\caption{Enhancement coefficient for $h/b$=1 (full line) and 8 (short-dashed line) diffraction gratings, illuminated by a normally incident light with electric field directed along the (a) $x$- and (b) $z$-axis, respectively.}
\label{Fig_3}
\end{figure}

In order to analyse the results in figure \ref{Fig_2}, $\eta$ is calculated for light arriving to the grating along the normal direction and for two limiting cases of polarization states, namely, those with the electric field perpendicular and parallel to the grooves, respectively. The results are plotted in figure \ref{Fig_3}, where, in order to avoid a busy plot, only the results corresponding to $h/b$=1 and 8 appear.   

The results in figure \ref{Fig_3}(a) clearly show that when the electric field is perpendicular to the grooves, the field enhancements are oscillating functions of $\lambda/b$, becoming scarcely larger than unity over nearly the entire range of wavelengths calculated in the present paper. On the contrary, when the electric field is directed along the $z$-axis, as shown in figure \ref{Fig_3}(b), $\eta$ exhibits a series of large peaks, regularly spaced, with an amplitude which appears to increase with increasing the wavelength. This is so, however, for $\lambda/b\leq$2, because for $\lambda/b>$2 the field enhancement falls below unity, and it does at a rate that seems to be an increasing function of the groove depth. This however is not at all unexpected, since $E_z$ must be zero all over the surface of the grating, therefore, a $E_z$-polarized light cannot penetrate within the groove as soon as its wavelength becomes comparable to, or larger than the groove width. 

\begin{figure}[ht]
\includegraphics[width=8.0cm,angle=-90]{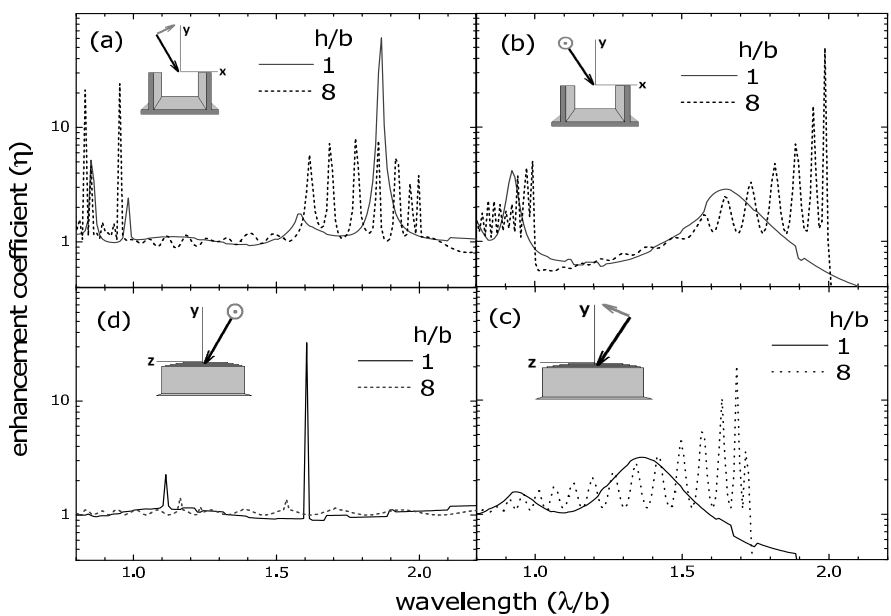}
\centering
\caption{Enhancement coefficient for $h/b$=1 (full line) and 8 (short-dashed line) diffraction gratings. Light arrives at 30 degree incidence angle ($\theta_i$) upon the grating, whereas the azimuthal angle ($\varphi_i$) and the electric field direction angle ($\phi$), both in degrees, are: (a) 0, 90; (b) 0, 0; (c) 90, 90; and (d) 90, 0. (see figure \ref{Fig_1} for an explanation of these angles) }
\label{Fig_4}
\end{figure}

Figure \ref{Fig_4} shows four limiting cases of beams arriving to the grating with 30 degree incidence angle ($\theta_i$). These comprise the TE-and TM-polarization state and, 0 and 90 degree azimuthal angles ($\varphi_i$). The first conclusion one may extract from these curves is that if the electric field of the light lies on a plane that is parallel to the grooves, such as the (b) and (c) cases, then, there appears a cut-off wavelength above which the field enhancement falls down below unity quite rapidly. However, if the electric field is perpendicular to the grooves, $\eta$ remains greater the unity over the entire range of wavelengths analysed in this paper. This is obvious, since no large electric fields can be developed within the grooves if they are parallel to the lateral and bottom surfaces. 

As was already observed elsewhere \cite{Glass83, Wirgin85, Llopis10, PIERS2011}, the results in figures \ref{Fig_4}(a-d) show a number of peaks, which can reach values as large as several decades. These peaks are observed to occur at wavelengths which depend on the groove depth $h$, the incidence and azimuth angles, and the state of polarization. Curiously though, such peaks are often so narrow that many of them nearly disappear after taking an average, as seen in figure 1.    

Finally, one may find the enhancement coefficient of the grating exposed to sunlight. In this case, one must take an average of $\eta$ over all polarisation states and wavelengths in the solar spectrum, 

\begin{equation}
\eta_{sun} = \frac{ \displaystyle \int_0^{\lambda_{0m}} d\lambda_0 \, \Phi_E(\lambda_0)\,\lambda_0\, \eta }{ \displaystyle\int_0^{\lambda_{0m}} d\lambda_0 \, \Phi_E(\lambda_0) \, \lambda_0\,   } \, ,
\label{Eq11}	
\end{equation}

\noindent where $\lambda_0$ is the wavelength of the light in vacuum, $\Phi_E(\lambda_0)$ is the spectral flux density of the sun light, $\eta$ is the enhancement coefficient for a given wavelength and after averaging over all polarisation states, and $\lambda_{0m}$ is the largest wavelength of the light which can promote electrons from the valence to the conduction band. Notice that the factor $\lambda_0$ appearing in both integrals is required in order to obtain the photon density spectrum from $\Phi_E(\lambda_0)$, which is obtained from the so-called Reference AM 1.5 spectra in \cite{AM1.5}. It must be also noticed that, since the wavelength in vacuum is used, the dimensions of the grating must be scaled using the refractive index $n$ of the medium that fills the groove. For the sake of simplicity, such an index is assumed to be constant along the solar spectrum and absorption is ignored, as a consequence  $\eta_{sun}$ will be a function of both the optical spacing of the grooves, i.e. $dn$, and the aspect ratio $h/b$.

\begin{figure}[h]
\includegraphics[width=10.0cm]{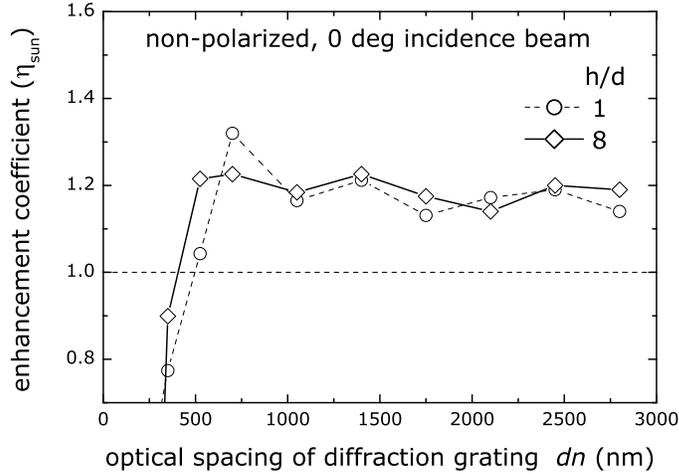}
\centering
\caption{Enhancement factor of the diffraction grating under AM 1.5 solar irrandiance assuming normal incidence. Calculations are performed for several values of the optical spacing of the grooves, i.e. $dn$, and two aspect ratios $h/d$=1 ($\bigcirc$) , and 8 ($\Diamond$).) }
\label{Fig_6}
\end{figure}

Equation (\ref{Eq11}) is calculated using the Monte Carlo method, and the results are plotted in figure \ref{Fig_6}. There, one can see that the enhancement coefficient becomes larger than unity for $dn$ approximately grater than 500 nm, whereas for $nd$ smaller than this value, $\eta_{sun}$ drops down fairly fast. For optical spacing between the grooves greater than 500 nm the enhancement coefficient becomes as large as 1.2  and stays around this value up to $dn$=3600 nm, which is the largest $dn$-value calculated in this paper. Remarkable though, in a similar fashion as was previously observed in figure \ref{Fig_2}, $\eta_{sun}$ appears to nearly not to depend on the aspect ratio $h/d$.

\section{Summary and concluding remarks}
\label{Summary}

The performance of a diffraction grating as a light trapping structure for PV cells applications is analysed. To this end, the electric fields produced within the grooves of a perfectly conducting, diffraction grating by a well collimated, monochromatic linearly-polarized beam of light are calculated. The grating has a $d$-period and is assumed to be made of infinitely long rectangular-shaped, $h$-deep and $b$-wide grooves. Although, for the purpose in this paper it is assumed that $d \approx b$. Furthermore, in order to properly assess the grating, an \emph{enhancement coefficient} ($\eta$) is introduced. $\eta$ is defined as the average electromagnetic-energy within the volume of the grooves relative to that one will have if grating is replaced by a flat perfect reflector. In this regard, a $\eta$ greater than unity implies that light is being trapped within the grooves of the diffraction grating. Results in this paper agree with previous calculations \cite{Glass83, Wirgin85, Lochbihler93, Depine94, Llopis10} that, for certain polarization state, wavelength and incidence angle, $\eta$ can be substantially larger than unity. When using conical, non-polarized beams, such enhancements however are reduced and $\eta$ can hardly be larger than three. Moreover, these values are observed for wavelength in the propagating medium within the range $b < \lambda < 2b$. For $\lambda < b$, $\eta$ appears to be slightly greater than unity irrespective of the polarization state and incidence angle, provided this later is not greater than 30 deg. For $\lambda > 2b$ however, the enhancement coefficient seems to be always smaller then unity. This is particularly so, when the electric field of the incoming light is parallel to the groove direction. Finally, the results of calculating the enhancement coefficient of the diffraction grating under the so-called Reference AM1.5 solar spectrum, namely $\eta_{sun}$, show that $\eta_{sun}$ can be, in the best of the cases, of the order of 1.2 . This maximum occurs for optical spacing of the groove $dn$ approximately equal to 500 nm, whereas for $dn\leq$ 500 nm $\eta_{sun}$ becomes negligible small. In all the cases, $\eta_{sun}$ does not seem to depend on the aspect ratio $h/b$. It is worth mentioning that, although the structure of the ideal cell sketched in this paper is different to that in Ref.\cite{Wang10}, the figures reported by these Authors are similar to those in the present calculations. In short, diffraction grating may conceivably act as a light trapping structure and, consequently, increase the efficiency of a PV cell, figures appear to be around 20\% for solar spectrum and groove spacing is expected to be in the sub-micron scale.

\subsection{Acknowledgements}
We are indebted to I. Tobias, A. Mart\'i and A. Luque for encouraging the Authors to work on this problem. This work has been supported in part by the European Union Program  IBPOWER (211640) and the NUMANCIA II project funded by the Comunidad Aut\'onoma de Madrid. 

\end{document}